*Snowmass White Paper*
*CMS Upgrade: Forward Lepton-Photon System*

Burak Bilki[2], Lucien Cremaldi[3], Yasar Onel[2], David R Winn[1*], Taylan Yetkin[2]

Fairfield University[1], University of Iowa[2], University of Mississippi[3]

*Correspondence: winn@fairfield.edu

*Abstract:*
This White Paper outlines a proposal for an upgraded forward region to extend CMS lepton (e, µ) and photon physics reach out to $2.2 \leq \eta \leq 5$ for LHC and SLHC, which also provides better performance for the existing or new forward hadron calorimetry for jet energy and $(\eta,\phi)$ measurements, especially under pileup/overlaps at high lumi, as LHC luminosity, energy and radiation damage increases.

*Introduction:* At the present time, the Forward Region $2.4<\eta<5.1$ has emerged as more favorable for new physics activity, in kinematics of signals, the need for high statistics, and fuller understanding of underlying events and the initial parton distribution functions (PDFs) over the x and $Q^2$ space. The main ideas motivating renewed interest in the forward region are:

*a) Light Higgs:* a relatively light (125 GeV) particle with properties highly consistent with the SM Higgs is produced over a larger range of $\eta$ and thus can be better studied with increased $\eta$ lepton coverage, including the 4 lepton final states, spin-parity measurements, direct H->µµ, and the investigation of unitary in VV scattering.

*b) SUSY:* the absence of a significant fraction of predicted SUSY particles at ~sub-TeV masses and resulting renewed emphasis on relatively light stops or a few other SUSY candidates and so missing $E_T$ becomes more crucial; as an example, a 1 TeV muon at $\eta=3$ has ~100 GeV ET.

*c) DM:* indications of relatively light dark matter, $\leq 10$ GeV, in underground experiments (annual variations in DAMA/LIBRA and CoGeNT, and recoil events in CRESST) indicate that lowering the missing ET threshold is crucial (see b) above).

*d) $B_d ->K^*\mu\mu, B_s -> \mu\mu$:* the dimuon decays of the neutral B as a test of physics beyond the standard model, including SUSY. Recent results from LHCb and CMS strongly constrain some of SUSY parameter space. CMS covers $|\eta|<2$, while LHCb covers $2< \eta <4.5$. Extending CMS coverage would both enable a far larger sample set, and also a more direct comparison with LHCb for combining results.

*e) $A_{FB}$: Muon Pair Front/Back Asymmetries*: $A_{FB}$ tests V-A in the SM; deviations indicate new physics processes BSM.

*Forward ($|\eta|>2.3$) Physics Processes*
Physics benefitting from Forward Leptons & photons and an improved Forward HCal include:

**1) Muon Pair F/B Asymmetries** - $A_{FB} = \frac{\sigma_F - \sigma_B}{\sigma_F + \sigma_B}$ where $\sigma_{F,B}$ is the differential angular cross-section of $qq \to \mu\mu$, integrated over each respective hemisphere. If we add more µ's from $2.4<\eta<5.1$ to CMS, we collect 50% more events, but with a x2 smaller fitted error, and a 4% larger $A_{FB}$[1]. $A_{FB}$ tests V-A in the SM; deviations indicate new physics processes BSM.

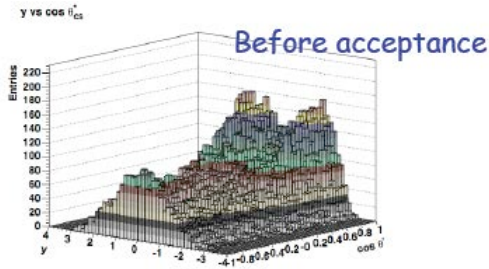

*Figure 1:* The number of asymmetry events vs y and cosθ* before acceptance cuts, with the table for fitted events, and the asymmetry variables.

*2) Low Mass Higgs Properties: Production, Decays, and Spin-Parity*: A 125 GeV (Standard Model) Higgs makes the acceptance for leptons and jets (VBF jet tags) in the forward region more important than for much higher mass Higgs. The channel H-> ZZ* -> 4 leptons results in about 40% of the Higgs decays throwing *at least* one lepton into the region 2.4<$\eta$<5, as shown below. Being able to accumulate high precision statistics on any of the Higgs decays is important to constrain physics beyond the Standard Model

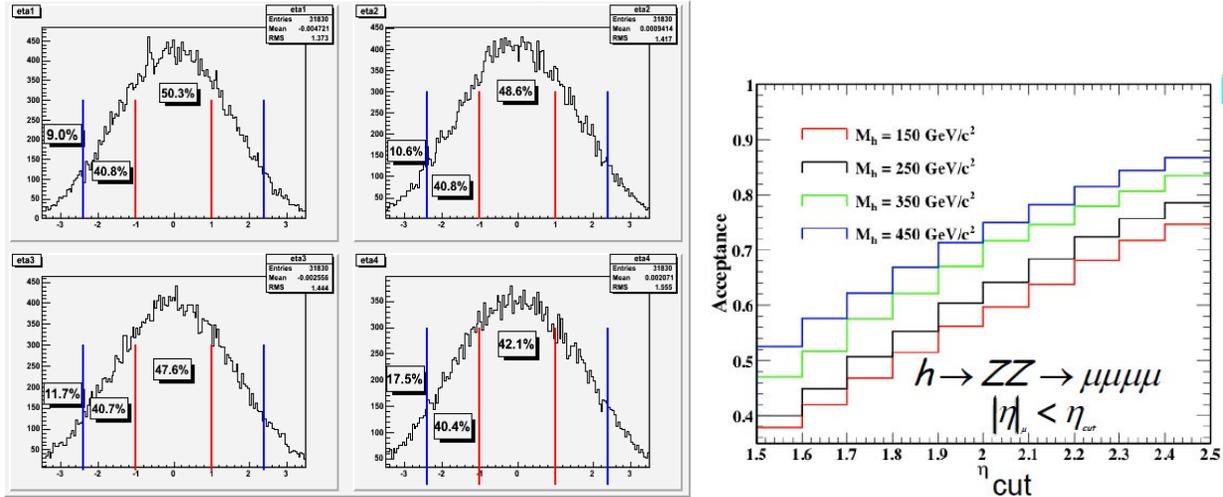

*Figure 2a:* The η-distribution of the each of the 4 leptons H-> ZZ -> $l_1 l_2 l_3 l_4$. About 35-40% of the Higgs decays to 4l have at least 1 lepton at η>2.4.
*Figure 2b*: The efficiency of H-> ZZ -> $l_1 l_2 l_3 l_4$ for 4 Higgs masses vs η cut. Extrapolating to mHiggs = 125 GeV, we estimate about 35% more Higgs would be accepted if η cut = 5.2

Moreover, Higgs *direct decays to muons* ie H–>μμ is about 0.25% of the BR of a 125 GeV Higgs, but requires >100 fB$^{-1}$; this important channel – the width of the Higgs and the absolute cross-section can be measured well - but is only accessible at very high lumi, and thus benefits from the nearly 20% of such muons lost without a forward lepton system.

Finally, the 4 lepton Higgs events can be used to separate the spin-parity of the candidate Higgs.

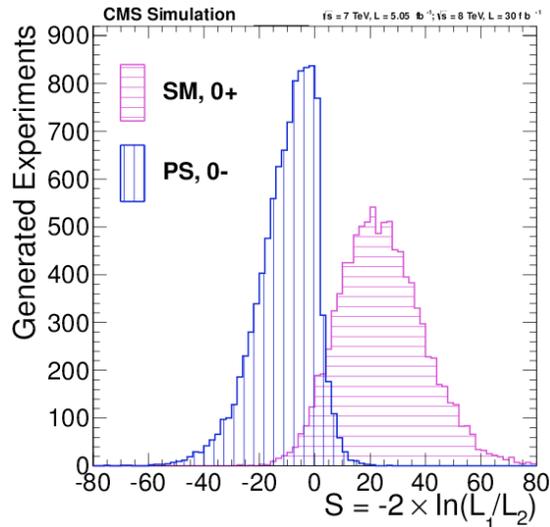

*Figure 2c*: A MELA analysis of $J^{PC}$ of the Higgs from leton angular measurements of the decay H-> ZZ -> $l_1l_2l_3l_4$ – a nearly 3σ separation at 30 fb$^{-1}$ using only muons from the central region – extending to higher eta may gain as much as 40% of the events (CMS Simulation.)

*3) Vector Boson Fusion*: VBF/Scattering is an important process for Higgs production and other heavier objects, requiring high statistics for both the jet and leptonic decays of WW, WZ, ZZ. Since color is not exchanged between the colliding protons, forward jets are preferentially produced, and forward jet tagging enhances VBF signals, as shown in the 160 GeV Higgs example below. The highest S/N is for jets at η>2.5 (about 3.1), and a rapidity difference between tagging jets $\Delta\eta_{jj}$ > 5 for cleanest signal/noise. The lighter 125 GeV Higgs enhances activity in the forward region.

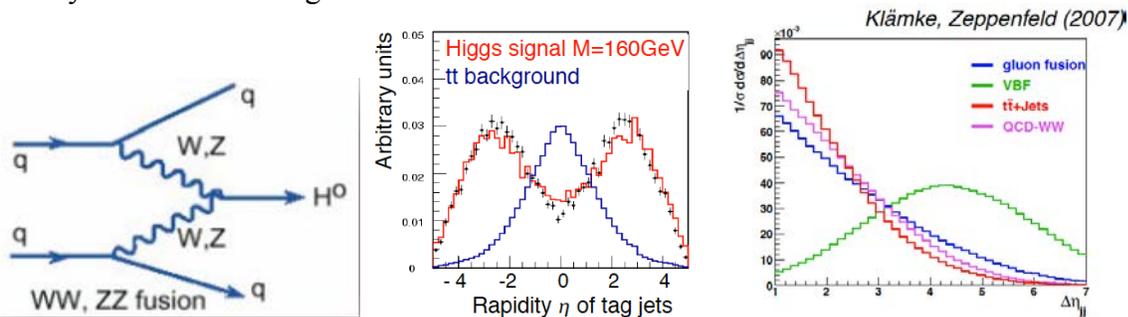

*Figures 3a-c:* VBF signal and background showing forward tagging jets for $m_H$ = 160 GeV. For the 125 GeV Higgs, the forward jets will be a somewhat higher rapidities.

Information on the CP properties of the Higgs produced by VBF can be obtained from the azimuthal angular difference $\Delta\phi_{jj}$ of the 2 tagging jets as shown in the figure below. The dip structures at 0/180° or 90° depend only on the tensor structure of the HVV coupling.

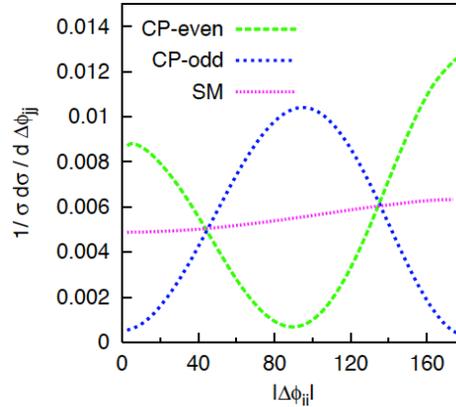

*Figure 4:* CP structure indicated by the azimuthal angle between the 2 tagging jets, requiring high lumi and analysis during pileup[2]

*4) EW Symmetry Breaking (EWSB) and Vector-Vector Scattering:* It is essential to show that the s-wave scattering amplitude of WW, WZ, ZZ is damped by the Higgs, protecting unitarity. Because the scattering is forward peaked in the light Higgs case, about 15% of the leptons from H-> WW -> $l_1 l_2$ have one of the leptons within the tagging jet cone. For a 125 GeV Higgs, the cross-section for $\Delta\phi_{ll}$ of the 2 leptons > 170° is ~50% higher than the no-Higgs case. These studies require greater than ~500 fb$^{-1}$ and so operation of the HF will require better jet definition a provided by a forward Ecal.

*5) SUSY:* If the LSP is <50 GeV, then missing energy cuts must be lowered. As an example of holes in the detector, a 500 GeV muon at η=3 carries away ~50 GeV in $E_T$, and at present does even carry a muon ID tag. At 14 TeV, forward muons and mismeasured jets carrying away $E_T$ is important. Similarly, SUSY searches for signals with >4 jets are increasingly important, while the lumi and hence overlapping events are increasing; jet ID and isolation become increasingly important. An e-m calorimeter with finer transverse segmentation will provide tighter $\Delta(\eta,\phi)$ jet definition.

*6) $B_d$->$K^*\mu\mu$, $B_s$ -> $\mu\mu$:* The dimuon decays of the neutral B test physics beyond the standard model, including SUSY. Recent results from LHCb and CMS strongly constrain some of SUSY parameter space. CMS covers |η|<2, while LHCb covers 2< η <4.5. Extending CMS muon coverage would both enable a far larger sample set, and also a more direct comparison with LHCb for combining results. A measurement would require tracker information sufficient to find 0.5 mm displaced B-decay secondary vertices.

*7) Standard Model Z, W Production (QCD-EW) and PDF's* - In general, Z or W (QCD-EW) production results in about 15-20% of the vector boson decay leptons to fall into the region 2.4<η<5, and ~10-12% fall into 3<η<5. Measuring W, Z and other Drell-Yan light processes, such as low pT resonance production of J/Ψ and Υ out to the highest η are sensitive to measuring and constraining the PDF's at low x, important for unfolding most of the production processes, as CMS transitions to 14 TeV.

***8) Double Parton Scattering:*** The correlation in x amongst quarks/partons is interesting; if we knew $F_2(x_1, x_2,..x_n)$, and Fourier transformed it, we would have the proton. Double Drell Yan is an avenue towards that but requires high statistics and acceptance for pairs of lepton pairs.

***9) Single Top QCD-EW Production:*** t-channel single top production (t-bbar/b-tbar production via W+b+gluon in the t-channel) is about ¼ of all ttbar production, but is tagged by a single forward jet; for tag-jets at rapidities $\eta>3$, the S:B exceeds 2.5:1, and growing with $\eta$. An important issue for such jets is overlapping events, which the present HF calorimeter is not ideal due to transverse segmentation and energy resolution. An e-m front end, finely segmented, would supply information leading to refined jet-cones.

***10) Exotica:……*** Heavy resonances, Z'/W', heavy quasi-stable charged particle precision timing. Both forward leptons and jets increase the physics reach and refine rejection or tagging processes.

*The Present Forward Situation:*

Beyond $\eta=2.6$, only hadron calorimeter measurements exist in CMS, without redundancy - no tracker, e-m calorimeter nor muon system for redundancy, only the full fiber and 1 lambda pulled back fibers of the quartz fiber forward calorimeter's non-independent "compartments". The HE extends to $\eta=3$ where HF takes over $5<\eta<3$, with CASTOR beyond. In CMS, there are a vigorous programs to upgrade the raddam resistance of the highest parts of HE out to $\eta\sim3$, to extend the raddam hardness of the electromagnetic calorimeter, to upgrade the tracker, and to extend muon measurements out to $\eta\sim3$.

We propose considering as well an upgrade to completely cover leptons out to $\eta\sim5$ and improve Jet measurements for $5<\eta<3$. The poly plug in front of HF, the HF strong-back and outer shielding, the collar, and the rotating shield upstream of HF are completely passive, and present an opportunity to be replaced with active devices. More space from a possible TOTEM replacement is also possible, and integration with or replacement of CASTOR and upstream. We show the existing region below.

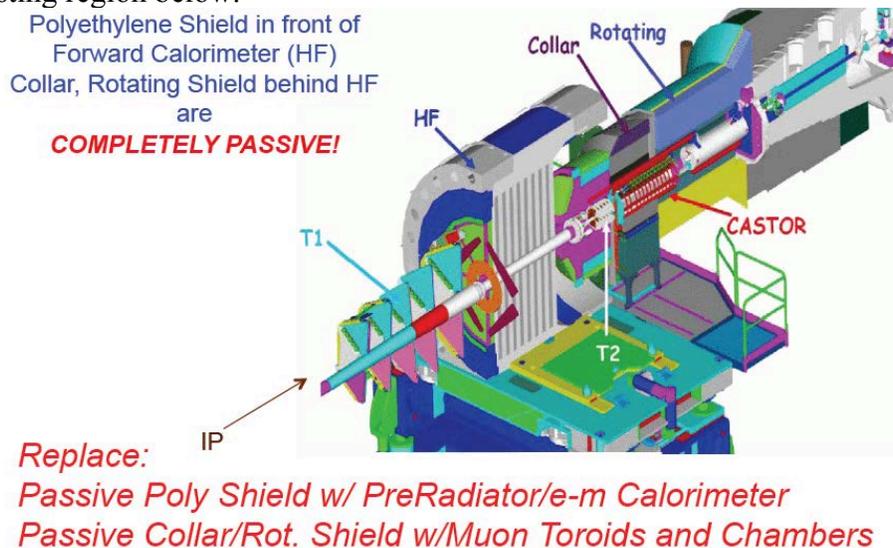

*Figure 5:* Passive regions that could be replaced with active devices: poly plug in front of HF, the HF strong-back and outer shielding, the collar, and the rotating shield upstream of HF are completely passive.

*The Present CMS Forward Calorimeter HF:*
The HF as designed is sufficient for forward jet measurements with the tower transverse size set approximately at the jet cone size which makes pileup and overlapping jets separation an issue in future at CMS at LHC and SLHC, as there is no tracker or e-m cal to provide finer segmentation or aid in jet definitions. It integrates over less than one crossing, a major plus. Its other deficit is that it is too thin at ~9 lambda, so punch-through to PMT's becomes problematic, with low energy resolution and signal size, and also to a possible muon system. The resolution is also compromised by the large e/pi response, and the non-independence of the long-short compartments, as well as a low response 0.5 pe./GeV even after the Stage 0 Upgrade. The recent Stage 0 upgrade of more sensitive multianode PMT viewing the same tower lessens some of the latter problems. We would like to concentrate in this White Paper on Lepton-Photon systems, assuming modest changes to HF for now.

A new replaceable stub tracker, preshower and e-m front end, absorbing most of the electromagnetic portion of the jets could: a) protect the front of the HF from raddam; b) define the jet direction and separation between forward jets more precisely; c) provide better energy resolution; d) help reduce punch-through to the HF readout. A new muon system behind HF would be able to identify punch-through from HF.

However, long term upgrades to the HF should be considered in tandem with a forward lepton-photon system. If the present HF quartz fiber Cerenkov metal matrix is not wholly replaced, besides fiber replacement, these include: 1) much higher transverse segmentation for pileup/jet overlap for LHC at full energy & lumi, and beyond; 2) photodetectors with more sensitivity and less punchthrough sensitivity; 3) a thicker HF, effectively 12 interaction lengths; 4) better electromagnetic energy response. The latter 2 issues can be part from an e-m front end of 1-1.5 Lint, and a rear compartment of 1-1.5 Lint, as the start of a muon system.

A fairly straight-forward replacement of or modification of the existing HF (if allowed by the activation levels) is possible with performance equal to SLHC tasks. Entirely new technology is also possible. We will describe possible changes in the existing HF or a new HF entirely in separate White Papers. Briefly, one such possibility is a haircut. The quartz fibers could be terminated at the rear face of the HF, and an array of highly-redundant photodetectors, with relatively low signals from punchthrough particles, with *at least* 3x3 times finer transverse segmentation would be mounted on the back face, as replaceable cassettes. Such detectors could be APD of Si, GaAs, SiC or others, or multianode metal envelope PMT with 0.5mm alumina windows (alumina windows are <$2/cm$^2$ ) or SIPM-like sub-detectors. The present fiber bundle region would be replaced by a tail-catcher compartment of 1.5-2.5 lambda.

*A Forward Lepton-Photon System* –
We show cartoons of the space in the forward regions of the detector where a stub-tracker/pre-radiator, e-m calorimeter, and an superferric iron-toroid muon system could be added to the forward region. Careful consideration would need to be given to services, new muon forward tracker upgrade proposals, modifications or replacement of HF, changes to Castor and Totem, and issues with the LHC services. More extensive modifications could replace the Forward Hadron Calorimeter with a magnetized iron matrix for quartz fibers, and/or fabricate the HF top shielding and HF strongback with muon toroids to cover muons from $\eta$~2 to 4.5.

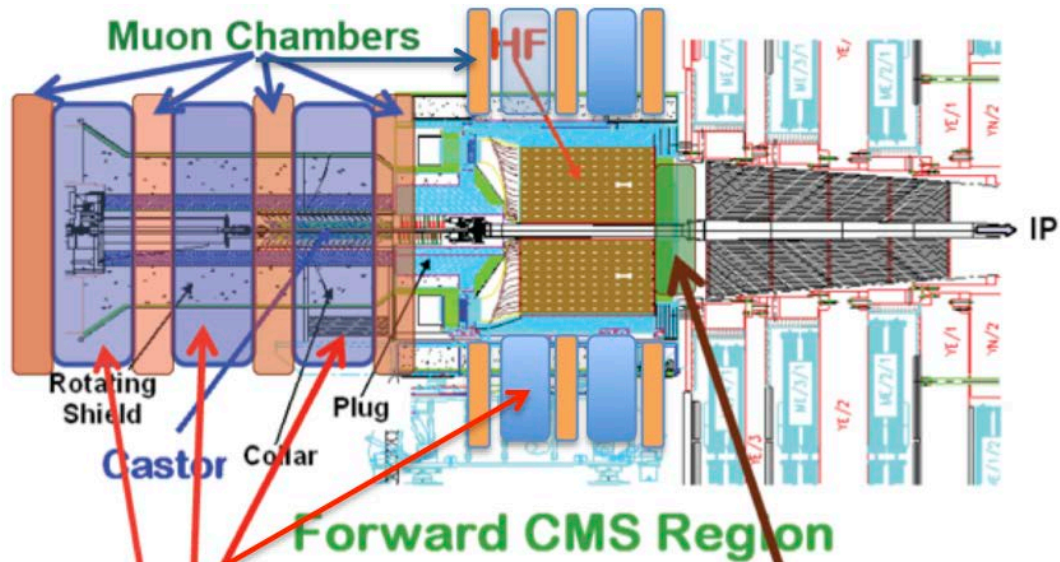

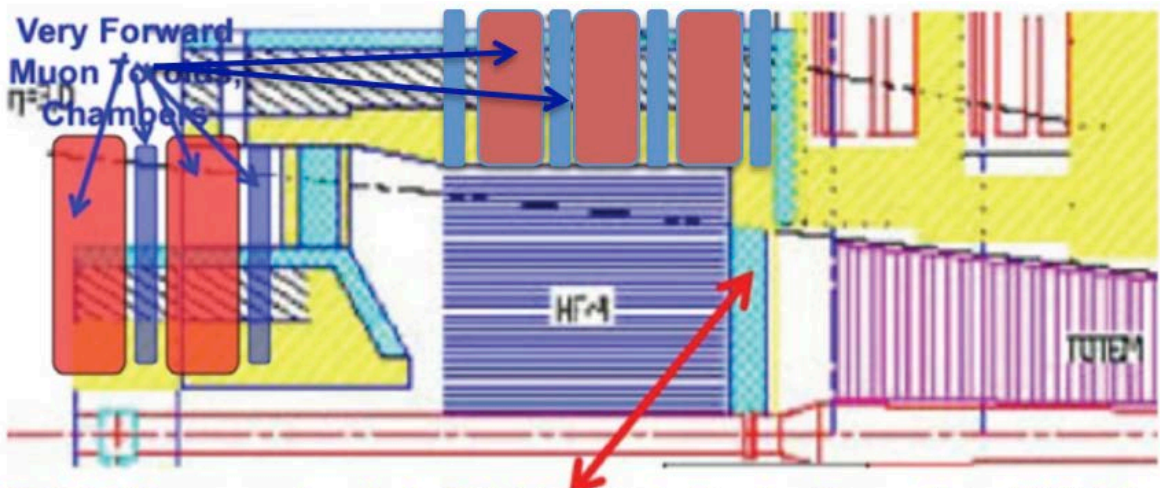

*Figures 6:* Cartoons (not fully to scales) of Lepton-Photon systems for the Forward Region: Passive regions that could be replaced with active devices:
1) poly plug in front of HF replaced by stub tracker, pre-radiator and E-M compartment;
2) the HF strong-back and outer shielding, the collar, and the rotating shield upstream of HF are replaced by superferric muon systems.

*Forward Muon System Example:*
SuperFerric Iron Toroids[3] are installed behind and also surrounding HF, with a total iron thickness ~3m. The toroids behind HF have an outer a radius of ~1.5 m and an inner radius as close to the beam pipe as technically feasible, the total thickness of Fe divided into 0.5-0.75m thick sections following HF. The toroids surrounding HF and an inner radius ~ 1.2m and an outer radius ~2.1m. These are interspersed with ~0.4m thick-along-z track measuring systems capable of +/- 0.05 mrad resolution per stack (~25 μm x-y resolution "chambers"), with a major radial dimension~1.4 m – (ACHTUNG! recycling Si tracker planes is interesting). The table below outlines the parameters. A resolution $\sigma_p/p$ < 15% at 1.5 TeV seems plausible. It may prove useful to have ionization information in the first planes of track-measurement to tag hard muon scatters.

- 3 m Fe Toroids: 4x0.75m Segments (Or more!)
- ~1.8 T Saturation minimum:
  - Superferric $H_{TC}LN_2$ –2.6T inner R, 2.1 T at R=1.3 m
- ~1-1.4 m max radius chambers, assume 20-50 μm
  - *Recycle The Tracker after Phase II!*
- 0.5 m wide chamber stacks (+/- 0.05 mrad per stack)
- $\sigma_p/p$ ~11% MS limit @ 1.5 TeV (4 Toroids/3m total of super-Fe)
  - p term in dp ($B^2L$) ~$10^{-4}$ p (GeV)

Possible further places for Fe toroids include the outer radius region of HFcal – replacing the strong back and shielding, at ~ 2.2≤ η ≤3, with HF inside the toroids.

*A Forward Stub-Tracker:*
A stub tracker is used to identify charged particles entering the Forward System and measure their x-y coordinates, with a spatial resolution of ~mm. A ~1 mm pixel at 11 m is nearly the same angular resolution as a 100μm pixel at 1.1 m. There are many technological choices but all must work at speed, pileup and high dose. Candidates include systems based on:
a) Non-polymerizing gas amplification, such as micromegas, MPGD/triple GEM;
b) Quartz fibers or bars;
c) Systems based on rad-hard pixels as APD in diamond, SiC, or forms of GaAs;
d) Systems based on secondary emission, such as dynode arrays, or fine-pitch multianode PMT with quartz or sapphire windows. PMT have proven to be excellent direct particle detectors via the Cerenkov light in the window, as the forward calorimeter has demonstrated;
e) Others as may be proposed.

*A Forward Pre-Shower/Preradiator for the e-m compartment:*
A preradiator[4] is used to identify electron from photon-initiated showers in the e-m calorimeter which follows the pre-shower. It is therefore essential for electron measurements without magnetic field information. Together with the tracker, a preshower will help identify which photons (and charged particles) are part of jet cones and define jet cones and especially overlapping jets in the forward hadron calorimeter, while helping to identify which electrons are incident and could be from prompt decays. The preshower would also give some modest angular information to charged hadrons incident. The pre-shower would use the same technologies as stub tracker, and consist of 4-6 planes separated by 0.25 Lrad radiators of W foils for compactness.

*A Forward Electromagnetic Calorimeter:*

A replaceable stub tracker, preshower and e-m front end, absorbing most of the electromagnetic portion of the jets could:
a) protect the front of the HF from raddam;
b) define the jet direction and separation between forward jets more precisely; c) provide better energy resolution;
d) help reduce punch-through to the HF readout. The present specs for the em calorimeter are energy resolution of at least as good as $10\%/\sqrt{E}$, ~1-1.5 lambda and 22+ radiation lengths thick, transversely segmented at half the shower width, and without need for longitudinal segmentation readout.

The space for an em calorimeter is where the 30cm thick passive poly shield in front of HF now resides, and possibly 10-20 cm further into the region of Totem. The technologies for this include:
a) Parallel quartz fibers, rods or bars in a dense, high Z matrix such as W or Pb, with 33-50% quartz packing;
b) Plates of quartz, or quartz cladded with rad-hard scintillator.
c) Germania - solid bars of $GeO_2$, the analog of quartz, but with a higher Z and density, and transmission in the red and near IR;
d) Gas-based planes with high-Z absorbers as in the tracker and pre-radiator;
e) LXe with a clearing field in high-Z tubes of few mm diameter - scintillation decay ~2.2 ns;
f) Others as proposed by the collaboration. At present, several point designs have detectors of 15-20 mm thick per Lrad readout.

*SUMMARY:*

A Forward Lepton-Photon System for CMS is outlined, replacing passive components in the forward calorimeter and muon regions of CMS $\sim 2 \leq \eta \leq 5$. It consists of super-ferric iron muon toroids interspersed with high resolution track-vector chamber modules, both behind the forward calorimeter(HF) and surrounding the forward calorimeter. It could provide ~10% momentum resolution at 1 TeV. An electromagnetic front end of about 1 Lint thickness would cover $3 \leq \eta \leq 5$, and consist of a stub tracker of ~1mm resolution(~100µradian), a 2-3 Lrad pre-shower, and a 25 Lrad E-M calorimeter, with a design goal of 5Grad radiation resistance and a stochastic term of $10\%/\sqrt{E}$, at present without electron sign discrimination. Such a system has interest for refining measurements of the Higgs, for unitary tests, for refining data on standard model processes, and for physics beyond the standard model.